\documentclass[aps,pra,twocolumn,nofootinbib,a4paper,floatfix,10pt]{revtex4-1}

\usepackage{color}
\usepackage{bm}
\usepackage{amssymb,amsmath,color,graphicx}
\usepackage{amsfonts}
\usepackage{dsfont}
\usepackage{bbold}
\usepackage{graphicx}
\usepackage{float}
\usepackage{subfigure}
\usepackage{verbatim}
\usepackage{amsfonts}
\usepackage{dcolumn}
\usepackage{bm}
\usepackage{color}
\usepackage[dvipsnames]{xcolor}
\usepackage{listings}
\definecolor{zaffre}{rgb}{0.0, 0.08, 0.66}
\usepackage[colorlinks=true,urlcolor=zaffre,citecolor=zaffre,linkcolor=zaffre]{hyperref}
\usepackage{mathrsfs}
\usepackage{filecontents}

\makeatletter
\def\maketag@@@#1{\hbox{\m@th\normalfont\normalsize#1}}  % enumeration size normal
\makeatother

\begin{document}

\title{Finite-size effects in continuous-variable QKD with Gaussian post-selection}

\author{Nedasadat Hosseinidehaj$^1$}  \email{n.hosseinidehaj@uq.edu.au}

\author{Andrew M. Lance$^{2}$}  %\email{nwalk@zedat.fu-berlin.de}

\author{Thomas Symul$^{2}$}  %\email{nwalk@zedat.fu-berlin.de}

\author{Nathan Walk$^{3}$}  %\email{nwalk@zedat.fu-berlin.de}

\author{Timothy C. Ralph$^1$}  %\email{ralph@physics.uq.edu.au}

\affiliation{$^1$Centre for Quantum Computation and Communication Technology, School of Mathematics and Physics, University of Queensland, St Lucia, Queensland 4072, Australia.}

\affiliation{$^2$QuintessenceLabs Pty. Ltd., Canberra ACT, Australia,}

\affiliation{$^3$Dahlem Center for Complex Quantum Systems, Freie Universit{\"a}t Berlin, 14195 Berlin, Germany.}

\date{\today}

\begin{abstract}

In a continuous-variable quantum key distribution (CV-QKD) protocol, which is based on heterodyne detection at the receiver, the application of a noiseless linear amplifier (NLA) on the received signal before the detection can be emulated by the post-selection of the detection outcome. Such a post-selection, which is also called a measurement-based NLA, requires a cut-off to produce a normalisable filter function. Increasing the cut-off with respect to the received signals results in a more faithful emulation of the NLA and nearly Gaussian output statistics at the cost of discarding more data. While recent works have shown the benefits of post-selection via an asymptotic security analysis, we undertake the first investigation of such a post-selection utilising a composable security proof in the realistic finite-size regime, where this trade-off is extremely relevant. We show that this form of post-selection can improve the secure range of a CV-QKD over lossy thermal channels if the finite block size is sufficiently large and that the optimal value for the filter cut-off is typically in the non-Gaussian regime. The relatively modest improvement in the finite-size regime as compared to the asymptotic case highlights the need for new tools to prove the security of non-Gaussian cryptographic protocols. These results also represent a quantitative assessment of a measurement-based NLA with an entangled-state input in both the Gaussian and non-Gaussian regime.

\end{abstract}

\maketitle

\section{Introduction}

Quantum key distribution (QKD) \cite{Scarani.et.al.RVP.09,Xu.et.al.arxiv.19,Pirandola.et.al.arxiv.19} is the most mature application of quantum information technologies, which allows two distant trusted parties, traditionally called Alice and Bob, to share a secret key which is unknown to a potential eavesdropper, Eve. In the quantum communication part of QKD Alice encodes classical information (i.e., key information) into conjugate quantum basis states, which are then transmitted over an insecure quantum channel to Bob, who measures the received quantum states in a randomly-chosen basis, to obtain classical information which is correlated to Alice's data. Repeating this procedure many times, Alice and Bob end up with two sets of correlated data, known as the raw keys. In the classical post-processing part of QKD Alice and Bob proceed with the sifting (if applicable), parameter estimation, reconciliation (or error correction), and privacy amplification over a public but authenticated classical channel to obtain a shared secret key \cite{Scarani.et.al.RVP.09,Xu.et.al.arxiv.19,Pirandola.et.al.arxiv.19}. QKD systems were first introduced using discrete-variable quantum systems, where the key information is encoded onto the degrees of freedom of single photons, and the measurement at the receiver is realized by single-photon detectors \cite{Bennett.Brassard.IEEE.84,Ekert.PRL.91}. As an alternative, continuous-variable (CV) QKD systems were introduced \cite{Ralph.PRA.99,Hillery.PRA.00,Reid.PRA.00}, where the key information is encoded onto the amplitude and phase quadratures of the quantized electromagnetic field, and the measurement relies on coherent detection, either homodyne or heterodyne detectors \cite{Garcia-Patron.PhD.07, Weedbrook.et.al.RVP.12, Diamanti}, which are faster and more efficient than single-photon detectors. CV-QKD systems can potentially achieve higher secret key rates than their discrete-variable counterparts, and their practical implementation is also compatible with current telecommunication optical networks. Although, thanks to the reverse reconciliation \cite{RR2003} (where the receiver, i.e., Bob, is the reference of the error correction), a secret key can asymptotically be generated for a pure loss channel over an arbitrary large distance, the practical secure distance of CV-QKD systems is limited due to the excess noise, imperfect classical post-processing, and finite-size effects.

In order to improve the transmission range of CV-QKD systems a post-selection strategy was proposed \cite{Silberhorn-PRL-2002, Lance-PRL-2005, Lorenz-ApplPhysB-2004, Lorenz-PRA-2006, Heid-PRA-2006}, in which, following the measurement of all the received quantum states, Alice and Bob
%decide, based on a pre-established rule, whether they should discard or keep parts of their classical data. In fact, Alice and Bob
discard the classical data corresponding to those channels for which the resulting key rate is negative, keeping only the data corresponding to those channels with a positive key-rate contribution. In this technique the resulting post-selected data has non-Gaussian statistics.
Further, it has been shown that the application of a noiseless linear amplifier (NLA), proposed in \cite{Ralph-QCMC-2009}, on the received signal preceding the detction can probabilistically enhance the secure range of CV-QKD systems, while preserving the Gaussian statistics  \cite{Blandino-PRA-2012}.
%NLA can amplify probabilistically the amplitude of an input state while preserving its noise characteristics \cite{Ralph-QCMC-2009}.
 However, any physical realization of the NLA is very demanding, requiring state-of-the art technology, such as single-photon sources, single-photon addition and subtraction. Moreover, the actual success probability of these experiments is much lower than the theoretical predictions. In \cite{Fiurasek-PRA-2012, Chrzanowski-NaturePhotonics-2014, Walk-PRA-2013} it has been shown that the physical implementation of the NLA can be substituted with a classical data post-processing. In particular, where the NLA directly precedes a heterodyne detection, the noiseless amplifier can be emulated by a Gaussian post-selectin of the detection outcome via a probabilistic classical filter function \cite{Fiurasek-PRA-2012, Chrzanowski-NaturePhotonics-2014}. This post-selection scheme, which is also called measurement-based NLA \cite{Chrzanowski-NaturePhotonics-2014}, has experimentally been demonstrated in \cite{Chrzanowski-NaturePhotonics-2014} and requires a cut-off on the classical filter to emulate the NLA.  The post-selection scheme results in Gaussian statistics for the post-selected data if the filter cut-off is chosen sufficiently large \cite{Fiurasek-PRA-2012, Chrzanowski-NaturePhotonics-2014}.

%However, the second technique proposed in \cite{Fiurasek-PRA-2012, Walk-PRA-2013} and experimentally demonstrated in \cite{Chrzanowski-NaturePhotonics-2014} is a Gaussian post-selection. The Gaussian post-selection of Bob's measurement outcome was introduced to emulate the noiseless linear amplification of the received signal before Bob's detection.

In the asymptotic regime it has been shown that the Gaussian post-selection can extend the maximum transmission distance of CV-QKD systems \cite{Fiurasek-PRA-2012, Chrzanowski-NaturePhotonics-2014, Walk-PRA-2013}. However, in reality a finite number of signals are exchanged between Alice and Bob.
The finite-size issue becomes even more significant when the post-selection is applied as it reduces the size of data.
%Hence, finite signalling is expected to have a large impact on the security analysis of CV-QKD systems with post-selection.
It is unclear whether the post-selection can still improve the CV-QKD performance in the realistic finite-size regime.

In this work we investigate the finite-size effects in the security analysis of CV-QKD systems with the post-selection (or the measurement-based NLA) at the receiver. We show that in the finite-size regime when the filter cut-off is large enough to make the post-selected statistics Gaussian, the maximum transmission distance of the CV-QKD system can be improved providing that the block size is sufficiently large. Considering finite blocks in a practical regime, we illustrate that the post-selection is effective when the CV-QKD system has undergone high values of excess noise.
Since reducing the cut-off can increase the success probability of the post-selection (at the expense of decreasing the classical mutual information between Alice and Bob), we also investigate the impact reducing the cut-off can have on the finite-size key rate, illustrating that if the filter cut-off is sufficiently reduced, the improvement of CV-QKD performance due to the post-selection can further be increased. Note that in the recent works on measurement-based NLA \cite{Fiurasek-PRA-2012, Chrzanowski-NaturePhotonics-2014}, the security proof is based on the equivalent entanglement-based scheme where the classical filter is replaced with a quantum filter (or NLA), as they assumed a sufficiently large cut-off to emulate an ideal NLA (with a Gaussian output). However, since reducing the cut-off can change the statistics of the post-selected data from Gaussian to non-Gaussian regime, we analyse the security proof based on the equivalent entanglement-based scheme with a classical filter (and not a quantum filter). Thus, our results also provides a characterization of the measurement-based NLA, when it is applied to a mixed Gaussian entangled state, which is an extension to a recent work on the characterization of the measurement-based NLA with a pure coherent-state input \cite{Zhao-PRA-2017}.

The structure of the remainder of this paper is as follows. In Sec.~II, the Gaussian CV-QKD system  is described. In Sec.~III, the security of the CV-QKD system is analysed in both the asymptotic and finite-size regime. In Sec.~IV, the post-selection of Bob's detection outcome is discussed, and the security of the post-selection protocol is analysed. In Sec.~V, the numerical results, showing the impact of the post-selection on the CV-QKD performance in the finite-size regime, is provided.  Finally, concluding remarks are provided in Sec.~VI.

\section{CV-QKD system}

Here we consider a Gaussian no-switching CV-QKD protocol \cite{no-switching1, Lance-PRL-2005}, that relies on the preparation of coherent states and heterodyne detection . In a prepare-and-measure scheme Alice generates a pair of random real numbers, $a_q$ and $a_p$, chosen from two independent Gaussian distributions of variance $V_A$. Alice prepares coherent states by modulating (displacing) a coherent laser source by amounts of $a_q$ and $a_p$, such that the variance of the imposed signals is $V_A$. The variance of the beam after the modulator is $V_A+1=V$ (where the 1 is for the shot noise variance), hence we obtain an average output state which is thermal of variance $V$. The prepared coherent states are then transmitted over an insecure quantum channel with transmissivity $T$ and excess noise $\xi$ (relative to the input of the quantum channel) to Bob. For each incoming state, Bob uses heterodyne detection and measures both the $\hat q$ and $\hat p$ quadratures for obtaining $(b_q, b_p)$. In this protocol, sifting is not needed, since both of the real random variables generated by Alice are used for the generation of the key. When the quantum communication is finished and all the incoming quantum states have been measured by Bob, classical post-processing including discretization, parameter estimation, error correction, and privacy amplification over a public but authenticated classical channel is commenced to produce a shared secret key.

This Gaussian CV-QKD system in the prepare-and-measure scheme can be represented by an equivalent entanglement-based scheme \cite{Garcia-Patron.PhD.07,Weedbrook.et.al.RVP.12}, where Alice generates a pure Gaussian entangled state, i.e., a two-mode squeezed vacuum state $\rho_{AB}$ with the quadrature variance $V$, where $V=\frac{1+\chi ^2}{1-\chi ^2}$, and where $\chi=\tanh(r)$, with $r$ being the two-mode squeezing parameter. Alice retains mode~$A$, while sending mode~$B$ to Bob. In the entanglement-based scheme, if Alice applies a heterodyne detection to mode~$A$, she projects mode~$B$ onto a coherent state. At the output of the channel, Bob applies a heterodyne detection to the received mode. As a result of Alice and Bob's heterodyne detection on all the shared entangled states, they end up with two sets of correlated classical data as the raw key, from which they can extract a shared secret key through the classical post-processing. % with his detector having an efficiency of $\eta$ and electronic noise variance of $\upsilon_{\rm el}$ \cite{inefficient_homodyne, inefficient_heterodyne}.

\section{Asymptotic and finite-size security analysis}\label{Asymptotic}

In the asymptotic regime the secret key rate in the reverse reconciliation scenario, where Bob is the reference of reconciliation, is given by $K = \beta I(a{:}b) - \chi(b{:}E)$ against Gaussian collective attacks, where $I(a{:}b)$ is the maximum mutual information shared between Alice and Bob limited by the Shannon bound, $\chi(b{:}E)$ is the maximum mutual information shared between Eve and Bob limited by the Holevo bound, and $0 \le \beta \le 1$ is the reconciliation efficiency. Note that in the asymptotic regime collective attacks are as strong as coherent attacks \cite{Renner2009, Garcia-Patron.PhD.07,Weedbrook.et.al.RVP.12}. Furthermore, for Gaussian CV-QKD protocols, where the key encoding is performed by a Gaussian  modulation of Gaussian states and the decoding is performed by Gaussian measurement, i.e., homodyne or heterodyne detection, Gaussian attacks are asymptotically optimal among collective attacks \cite{Gaussian-optimality-1, Gaussian-optimality-2, Gaussian-optimality-3}.

% \section{Finite-size security analysis}\label{Finite-size}

In the finite-size regime, the Gaussian no-switching CV-QKD protocol acting on $2n$ coherent states sent from Alice to Bob (or $2n$ two-mode squeezed vacuum states in the equivalent entanglement-based scheme) is $\epsilon$-secure against Gaussian collective attacks in the reverse reconciliation scenario if $\epsilon {=} 2\epsilon_{\rm sm}{+}\bar \epsilon {+}\epsilon_{\rm PE}{+}\epsilon_{\rm cor}$ \cite{Finite-size-Leverrier,Finite-size-Lupo} and if the key length $\ell$ is chosen such that \cite{Finite-size-Leverrier,Finite-size-Lupo}
\begin{equation}\label{key-length-main}
\begin{array}{l}
 \ell  {\le} N[ \beta I(a{:}b) {-} \chi(b{:}E) ] {-} \sqrt N {\Delta } {-} 2\log_2 (\frac{1}{{2\bar \epsilon }}),
 \end{array}
\end{equation}
where \cite{Finite-size-Leverrier,Finite-size-Lupo}
\begin{equation}\label{delta-AEP}
\begin{array}{l}
\Delta = (d{+}1)^2{+}4(d{+}1)\sqrt{\log_2({2{/}\epsilon_{\rm sm}^2})} {+}\\
\\
 2\log_2({2}{/}({\epsilon^2 \epsilon_{\rm sm}}))  {+} 4{\epsilon_{\rm sm}d}{/}{(\epsilon \sqrt N) },
 \end{array}
\end{equation}
and where $N{=}2n$, $d$ is the discretization parameter, $\epsilon_{\rm sm}$ is the smoothing parameter, and $\epsilon_{\rm cor}$ and $\epsilon_{\rm PE}$ are the maximum failure probabilities for the error correction and parameter estimation, respectively. %Note that for the $\epsilon$-security analysis of the same protocol against individual attacks we can still use Eq.~(\ref{key-length-main}), where $\chi(b{:}E)$ must be replaced by the Shannon mutual information between Eve and Bob, $I(b{:}E)$.

The final key rate where the key is $\epsilon$-secure against Gaussian collective attacks is given by $\ell/N$. Note that in Eq.~(\ref{key-length-main}) we have considered the same scenario as \cite{Finite-size-Leverrier}, where almost all the raw data can be utilized to distill the secret key (by performing the parameter estimation after the error correction\footnote{It has also been shown in \cite{PE-MDI-2018} that in CV-QKD the whole raw keys can be used for both parameter estimation and secret key generation, without compromising the security, and without any requirements of doing error correction before parameter estimation.}). However, if Alice and Bob are required to disclose a non-negligible number of data points of size $k$, during the parameter estimation, a classical data of size $N'=N-k$ is used for the key extraction. As a result, the final secure key rate is given by $\ell/N$, where $\ell$ is given by Eq.~(\ref{key-length-main}), but now $N$ in Eqs.~(\ref{key-length-main}) and (\ref{delta-AEP}) has to be replaced by $N'$.

Note that according to the approach introduced in \cite{Finite-size-Leverrier2017,Finite-size-Leverrier2019}, and numerically analysed in \cite{Neda-Nathan-Tim}, in order to analyse the composable finite-size security of the no-switching CV-QKD protocol against coherent attacks, the security of the protocol is first analysed against Gaussian collective attacks with a security parameter $\epsilon$ \cite{Finite-size-Leverrier}, and then by applying the Gaussian de Finetti reduction \cite{Finite-size-Leverrier2017} the security is obtained against coherent attacks with a polynomially larger security parameter $\tilde \epsilon$ \cite{Finite-size-Leverrier2017}, where the security loss due to the reduction from coherent attacks to collective attacks scales like $\mathit{O}(N^4)$ \cite{Finite-size-Leverrier2017}.

%and where $I(A{:}B)$ is the Shannon mutual information between Alice and Bob (calculated and provided in Appendix.~B), $\chi(B{:}E)$ is the Holevo mutual information between Eve and Bob (calculated and provided in Appendix.~B), $\beta$ is the reconciliation efficiency,

% $N{=}2n$ , and where $d$ is the discretization parameter, and $\epsilon_{\rm cor}$ and $\epsilon_{\rm PE}$ are the maximum failure probabilities for the error correction and parameter estimation, respectively (see Appendix.~A for more details). We have considered the same scenario as \cite{Finite-size-Leverrier,Finite-size-Lupo, PE-MDI-2018}, where almost all the raw data can be utilized to distill the secret key. Note that for the $\epsilon$-security analysis of the same protocol against individual attacks we can still use Eq.~(\ref{key-length-main}), where $\chi(B{:}E)$ must be replaced by the Shannon mutual information between Eve and Bob, $I(B{:}E)$ (calculated and provided in Appendix.~B).

\section{{Post-selection}}

\subsection{Noiseless linear amplifier (quantum filter)}

In contrast to classical optical channels, losses in quantum channels cannot be compensated for by usual deterministic phase-insensitive amplifiers, as the latter would inevitably introduce additional noise \cite{Caves-PRD-1982}, making the quantum channel insecure. To avoid this noise penalty, the idea of heralded noiseless linear amplifier (NLA) was proposed in \cite{Ralph-QCMC-2009}, which enables one to amplify probabilistically the amplitude of a coherent state without adding any extra noise. An NLA can be represented by the unbounded amplification operator $g^{\hat n}$ with the amplification gain $g > 1$ and the photon number operator $\hat n$, which realizes the following transformation on an input coherent state $\left| \alpha  \right\rangle $ \cite{Ralph-QCMC-2009},
\begin{eqnarray}\label{NLA-coherent}
{g^{\hat n}}\left| \alpha  \right\rangle  = \exp \left[ {\frac{1}{2}({g^2} - 1){{\left| \alpha  \right|}^2}} \right]\left| {g\alpha } \right\rangle.
\end{eqnarray}
For a Gaussian CV-QKD system it has been shown that the maximum transmission distance of the system can be increased by applying an NLA on the received mode preceding Bob's detection \cite{Blandino-PRA-2012}. Explicitly, in the equivalent entanglement-based scheme of the CV-QKD system Alice prepares a pure two-mode Gaussian entangled state, keeps one mode, while sending the second mode through an insecure quantum channel to Bob, who applies an NLA to noiselessly amplify the received mode, and distill the entanglement. Since the amplification is probabilistic, the successfully distilled entangled states are then used in an ordinary deterministic CV-QKD protocol, where Alice and Bob apply Gaussian measurements to their own shared modes.

An ideal NLA probabilistically converts a Gaussian state into another Gaussian state. The NLA distills the entanglement between Alice and Bob, hence effectively converts the initial channel into another channel with presumably higher associated performances. It has been shown in \cite{Blandino-PRA-2012} that for an entanglement-based scheme with an initial pure entangled state with the two-mode squeezing parameter of $\chi$, and a quantum channel with the transmissivity $T$ and the excess noise $\xi$, the covariance matrix of the output amplified state is equal to the covariance matrix of an equivalent system with a two-mode squeezing parameter $\chi_g$, sent through a channel of transmissivity $T_g$ and excess noise $\xi_g$, without using the NLA. These effective parameters are given by \cite{Blandino-PRA-2012}
\begin{eqnarray}\label{eff-parameters}
\begin{array}{l}
{\chi_g} = \chi \sqrt {\frac{{({g^2} - 1)(\xi  - 2)T - 2}}{{({g^2} - 1)\xi T - 2}}} \\
\\
{T_g} = \frac{{{g^2}T}}{{({g^2} - 1)T\left[ {\frac{1}{4}({g^2} - 1)(\xi  - 2)\xi T - \xi  + 1} \right] + 1}}\\
\\
{\xi _g} = \xi  - \frac{1}{2}({g^2} - 1)(\xi  - 2)\xi T.
\end{array}
\end{eqnarray}
These effective parameters can be interpreted as physical parameters of an equivalent system if they satisfy the constraints $0 \le \chi_g < 1$, $0 \le T_g \le 1$, and $\xi_g \ge 0$. Note that the first condition of Eq.~(\ref{eff-parameters}) is always satisfied if $\chi$ is below a limit value \cite{Blandino-PRA-2012}
\begin{equation}\label{lambda-limit}
0 \le {\chi _g} < 1 \Rightarrow 0 \le \chi  < {\left( {\sqrt {\frac{{({g^2} - 1)(\xi  - 2)T - 2}}{{({g^2} - 1)\xi T - 2}}} } \right)^{ - 1}}.
\end{equation}
Recall that Eq.~(\ref{eff-parameters}) can only be utilized to calculate the covariance matrix of the output amplified state of an NLA, when the NLA can be ideally implemented to preserve the Gaussianity of the input state.

The improvement of the performance of Gaussian CV-QKD systems using an ideal NLA has been discussed for different protocols and in different scenarios \cite{NLA-CVQKD-Entropy,NLA-CVQKD-LO, NLA-CVQKD-noisy}. However, in all of these works the success probability has been considered based on the theoretical predictions (which is much higher than the actual experimental success probability).
%the NLA operation has been approximated using the non-Gaussian quantum scissors operation, e investigate
Also, the use of quantum scissors as a practical candidate for an NLA has been investigated in CV-QKD systems \cite{NLA-CVQKD-no-switching, razavi1, razavi2,Malaney-QS}. Note that all of these works have focussed on the CV-QKD performance in the asymptotic regime, which is an unrealistic scenario.
%(consisting of an entanglement distribution protocol with an untrusted source, an entanglement swapping protocol with an untrusted relay \cite{}, CV-QKD protocols with noisy coherent states, ) using NLA has been discussed in \cite{}, where the protocols .

\subsection{Measurement-based NLA (classical filter)}

Since all optical implementations of NLA are extremely challenging, the method of Gaussian post-selection or measurement-based NLA was proposed \cite{Fiurasek-PRA-2012, Walk-PRA-2013}, and experimentally demonstrated \cite{Chrzanowski-NaturePhotonics-2014}, where the physical implementation of an NLA can be emulated with a suitable data processing. This represents a significant advantage as the difficulty of sophisticated physical operations can be moved from a hardware implementation to a software implementation. In particular, it has been shown in \cite{Fiurasek-PRA-2012, Chrzanowski-NaturePhotonics-2014}, when an NLA directly precedes a heterodyne detection, the NLA can be emulated by conditioning upon the heterodyne measurement outcome via a classical filter function. This means that in the no-switching CV-QKD system, the probabilistic noiseless amplification of the received signal before Bob's heterodyne detection can be emulated by the probabilistic post-selection of Bob's heterodyne measurement data \cite{Fiurasek-PRA-2012, Chrzanowski-NaturePhotonics-2014}.

%An ideal noiseless linear amplifier (NLA) can be represented by the unbounded amplification operator $g^{\hat n}$ with the amplification gain $g > 1$ and the photon number operator $\hat n$, which realizes the following transformation on an input coherent state $\left| \alpha  \right\rangle $ \cite{Ralph-Lund-2009}:
%\begin{eqnarray}\label{NLA-coherent}
%{g^{\hat n}}\left| \alpha  \right\rangle  = \exp \left[ {\frac{1}{2}({g^2} - 1){{\left| \alpha  \right|}^2}} \right]\left| {g\alpha } \right\rangle.
%\end{eqnarray}
Considering the input state of an NLA as $\rho_{\rm in}$, the Husimi $Q$-function of the amplified output state is given by
\begin{eqnarray}\label{amplified-coherent-Qfun}
\begin{array}{l}
Q_{\rm out}(\alpha ) = \frac{1}{\pi }\left\langle \alpha  \right|{g^{\hat n}}{\rho _{\rm in}}{g^{\hat n}}\left| \alpha  \right\rangle  = \\
\\
\exp \left[ {({g^2} - 1){{\left| \alpha  \right|}^2}} \right]\frac{1}{\pi }\left\langle {g\alpha } \right|{\rho _{\rm in}}\left| {g\alpha } \right\rangle.
\end{array}
\end{eqnarray}
Performing a change of variable, $\alpha_m = g \alpha$, we obtain
\begin{eqnarray}\label{amplified-coherent-Qfun_change_variable}
\begin{array}{l}
{Q_{\rm out}}({\alpha _m}) = \exp \left[ {(1 - \frac{1}{{{g^2}}}){{\left| {{\alpha _m}} \right|}^2}} \right]\frac{1}{\pi }\left\langle {{\alpha _m}} \right|{\rho _{\rm in}}\left| {{\alpha _m}} \right\rangle
\end{array}.
\end{eqnarray}
Having Eq.~(\ref{amplified-coherent-Qfun_change_variable}), we are able to determine the appropriate classical post-selection filter to approximate an ideal NLA prior to a heterodyne detection.

Let us assume in the entanglement-based representation of the no-switching CV-QKD protocol, Alice and Bob share a mixed Gaussian entangled state $\rho_{AB}$ (with a zero mean and covariance matrix ${\bf{M}}= [a{\bf{I}}_2,c{\bf{Z}};c{\bf{Z}},b{\bf{I}}_2]$ with ${\bf{I}}_2$ a  $2 {\times 2} $ identity matrix, and $\bf{Z} = \rm diag\left( {1, - 1} \right)$) before the detection. When Alice and Bob apply heterodyne detection to their own modes, obtaining the measurement values $\alpha_m$ and $\beta_m$ respectively, the joint probability distribution of the measurement outcomes is given by $Q_{\rm in}({\alpha _m},{\beta _m})$, which is in fact the Husimi $Q$-function of the mixed Gaussian entangled state $\rho_{AB}$. Note that the Husimi Q-function of a Gaussian two-mode state with a zero mean and covariance matrix ${\bf{M}}$ can be expressed as \cite{Fiurasek-PRA-2012},
\begin{equation}\label{input-Qfunc}
\begin{array}{l}
{Q_{\rm in}}({\alpha _m},{\beta _m}) = \frac{{\sqrt {\det ({\bf{\Gamma}})} }}{{{\pi ^2}}} \times \\
\\
\exp \left[ { -  {a'{{\left| {{\alpha _m}} \right|}^2} {-} b'{{\left| {{\beta _m}} \right|}^2} {-} 2c'\left| {{\alpha _m}} \right|\left| {{\beta _m}} \right|\cos ({\phi _\alpha } {+} {\phi _\beta })} } \right],
\end{array}
\end{equation}
where ${\bf{\Gamma}} = [a'{\bf{I}}_2,c'{\bf{Z}};c'{\bf{Z}},b'{\bf{I}}_2] = 2({\bf{M}} + {\bf{I}}_4)^{-1} $ with ${\bf{I}}_4$ a  $4 {\times} 4 $ identity matrix. Note that we have $\alpha_m = \left| {{\alpha _m}} \right| \exp(i \phi _\alpha )$ and $\beta_m = \left| {{\beta _m}} \right| \exp(i \phi _\beta )$.

Post-selection in the CV-QKD protocol is performed by filtering of the raw key (i.e., the measurement outcomes) based on the value of the quadrature amplitudes detected by Bob. In fact, Bob applies a probabilistic filter to his measurement outcomes, $\beta_m$, to realize the pre-factor, $\exp \left[ {(1 - \frac{1}{{{g^2}}}){{\left| {{\beta _m}} \right|}^2}} \right]$, in Eq.~(\ref{amplified-coherent-Qfun_change_variable}). Note that the filter is truncated by a real cut-off parameter $\gamma_c$ to make the filter probability convergent. The filter function is \cite{Fiurasek-PRA-2012, Chrzanowski-NaturePhotonics-2014, Zhao-PRA-2017}
\begin{equation}\label{filter}
F({\beta _m}) = \left\{ {\begin{array}{*{20}{c}}
{\exp \left[ {(1 - \frac{1}{{{g^2}}})\left( {{{\left| {{\beta _m}} \right|}^2} - \gamma _c^2} \right)} \right],\,\left| {{\beta _m}} \right| < {\gamma _c}}\\
{1,\,\,\,\,\,\,\,\,\,\,\,\,\,\,\,\,\,\,\,\,\,\,\,\,\,\,\,\,\,\,\,\,\,\,\,\,\,\,\,\,\,\,\,\,\,\,\,\,\,\,\,\,\,\,\,\,\,\,\,\,\,\,\,\,\,\,\,\,\,\,\,\,\,\,\left| {{\beta _m}} \right| \ge {\gamma _c}},
\end{array}} \right.
\end{equation}
where $\beta_m = b_q + ib_p$ is constructed from Bob's quadrature measurement outcomes $b_q$ and $b_p$, and the first piece of $F({\beta _m}) $ gives the acceptance probability, with which particular heterodyne measurement outcomes of Bob (outcomes with magnitude less than $\gamma_c$) are kept, while the others beyond the cut-off $\gamma_c$ are kept with unity probability.

Considering $N_{\rm ps}$ as the number of accepted data points which are kept by Bob, and $N$ is the total number of data points before the post-selection, the success probability of the post-selection is given by
\begin{equation}\label{succ-prob}
\begin{array}{l}
{P_s} = \frac{{{N_{\rm ps}}}}{N} = {\int}{\int} d^2{\alpha _m}{\int}{\int} d^2{\beta _m}\,\,F({\beta _m})Q_{\rm in}({\alpha _m},{\beta _m})=\\
\\
\int\limits_{0}^{2\pi } \int\limits_{0}^\infty  {d{\phi _\alpha }}  {d\left| {{\alpha _m}} \right|} \int\limits_{0}^{2\pi } \int\limits_{0}^{{\gamma _c}} {d{\phi _\beta }}  {d\left| {{\beta _m}} \right|} \exp\left[{ {(1 {-} \frac{1}{{{g^2}}})\left( {{{\left| {{\beta _m}} \right|}^2} {-} \gamma _c^2} \right)} }\right] \\
\\
\times
{Q_{\rm in}}({\alpha _m},{\beta _m}) \left| {{\alpha _m}} \right| \left| {{\beta _m}} \right|  \\
\\
 + \int\limits_{0}^{2\pi } \int\limits_{0}^\infty {d{\phi _\alpha }}   {d\left| {{\alpha _m}} \right|} \int\limits_{0}^{2\pi } \int\limits_{{\gamma _c}}^\infty {d{\phi _\beta }}   {d\left| {{\beta _m}} \right|} \,{Q_{\rm in}}({\alpha _m},{\beta _m}) \left| {{\alpha _m}} \right| \left| {{\beta _m}} \right|,
\end{array}
\end{equation}
%where $\alpha_m = \left| {{\alpha _m}} \right| \exp(i \phi _\alpha )$ and $\beta_m = \left| {{\beta _m}} \right| \exp(i \phi _\beta )$.
The final step to emulate an NLA is a linear rescaling on Bob's side that realizes $\beta_m=g \beta$. However, the rescaling is only applied to Bob's measurement outcomes with magnitude less than $\gamma_c$, while the others beyond the cut-off $\gamma_c$ are kept unaffected. The final joint probability distribution of the measurement outcomes after the rescaling is given by
%\begin{equation}\label{output-Qfunc}
%{Q_{\rm out}}(\alpha ,\beta ) = \frac{1}{{{P_s}}}\left\{ {\begin{array}{*{20}{c}}
%\begin{array}{l}
%{g^2}F(g\beta ){Q_{\rm in}}(\alpha ,g\beta ),\,\,\,\left| {g\beta } \right| < {\gamma _c}\\
%\\
%
%\end{array}\\
%{{Q_{\rm in}}(\alpha ,\beta ),\,\,\,\,\,\,\,\,\,\,\,\,\,\,\,\,\,\,\,\,\,\,\,\,\,\,\left| \beta  \right| \ge {\gamma _c}}.
%\end{array}} \right.
%\end{equation}
\begin{equation}\label{output-Qfunc}
\begin{array}{l}
{Q_{\rm out}}(\alpha ,\beta ) = \\
\\
\left\{ {\begin{array}{*{20}{c}}
\begin{array}{l}
\frac{{{g^2}}}{{{P_s}}}\exp \left[ {(1 {-} \frac{1}{{{g^2}}})\left( {{{\left| {{\beta _m}} \right|}^2} {-} \gamma _c^2} \right)} \right]{Q_{\rm in}}({\alpha _m},{\beta _m}),\,\left| {{\beta _m}} \right| {<} {\gamma _c}\\
\\
\end{array}\\
{\frac{1}{{{P_s}}}{Q_{\rm in}}({\alpha _m},{\beta _m}),\,\,\,\,\,\,\,\,\,\,\,\,\,\,\,\,\,\,\,\,\,\,\,\,\,\,\left| {{\beta _m}} \right| \ge {\gamma _c},}
\end{array}} \right.
\end{array}
\end{equation}
where $\beta_m{=}g \beta$ for $\left| {\beta_m } \right| {<} {\gamma _c}$, and $\beta_m{=}\beta$ for $\left| {\beta_m } \right| {\ge} {\gamma _c}$, while Alice's measurement outcomes do not need rescaling, i.e., we always have $\alpha_m{=}\alpha$.

Thus, in the post-selection, Bob first applies the filter function, $\exp \left[ {(1 - \frac{1}{{{g^2}}})\left( {{{\left| {{\beta _m}} \right|}^2} - \gamma _c^2} \right)} \right]$, to his measurement outcomes $\beta_m$ with magnitude less than $\gamma_c$, and then rescales his filtered outcomes such that $\beta_m=g\beta$, while his measurement outcomes beyond the cut-off $\gamma_c$ are kept unaffected with unit probability. In the CV-QKD protocol with the post-selection, for each measurement, Bob publicly reveals whether the outcome is kept or rejected, in order for Alice to keep or discard her corresponding measurement outcome. The filtered raw key of size $N_{\rm ps}$ is then treated as if it was the original raw key, which means the parameter estimation (to estimate the covariance matrix, ${\bf{M}}_{\rm ps}$, of the post-selected state shared between Alice and Bob in the equivalent entanglement-based scheme) should be performed on the post-selected data.
%Note that the success probability of the post-selection can also be given by
%\begin{equation}\label{succ-prob}
%{P_s} = \frac{{{N_{\rm ps}}}}{N} = \int\int d^2{\alpha }\int\int d^2{\beta }\,\, Q_{\rm out}({\alpha},{\beta }).
%\end{equation}

Having the final probability distribution of the post-selected data, $Q_{\rm out}({\alpha},{\beta })$, we are able to calculate the inferred covariance matrix of the amplified state before the heterodyne detection in the equivalent quantum-filter representation. The inferred covariance matrix ${\bf{M}}_{\rm ps} = [a_{\rm ps}{\bf{I}}_2,c_{\rm ps}{\bf{Z}};c_{\rm ps}{\bf{Z}},b_{\rm ps}{\bf{I}}_2]$ is given by
\begin{equation}\label{CM-postselected}
\begin{array}{l}
a_{\rm ps} = {\int}{\int} d^2{\alpha}{\int}{\int} d^2{\beta} \,\, ([2{\mathop{\rm Re}\nolimits} (\alpha )]^2-1){Q_{\rm out}}(\alpha ,\beta ),\\
\\
b_{\rm ps} = {\int}{\int} d^2{\alpha}{\int}{\int} d^2{\beta} \,\, ([2{\mathop{\rm Re}\nolimits} (\beta )]^2-1){Q_{\rm out}}(\alpha ,\beta ),\\
\\
c_{\rm ps} = {\int}{\int} d^2{\alpha}{\int}{\int} d^2{\beta} \,\, (4{\mathop{\rm Re}\nolimits} (\alpha ){\mathop{\rm Re}\nolimits} (\beta )){Q_{\rm out}}(\alpha ,\beta ).\\
\end{array}
\end{equation}
%Note that in the entanglement-based scheme, ${\bf{M}}_{\rm ps}$ is the inferred covariance matrix of the amplified state \emph{before} the heterodyne detection.
Note that in Eq.~(\ref{CM-postselected}) only the second moment of Alice and Bob's quadratures has been calculated to compute the elements of the covariance matrix of the amplified state, since the first moment of Alice and Bob's quadratures remain zero after the post-selection.

% The advantage of post-selection is that the correlation between Alice and Bob can be significantly higher on the filtered data. Under certain circumstances (discussed later) it can change an insecure protocol into a secure one. On the other hand, often a significant amount of data needs to be sacrificed in the filtering stage.

\subsection{Security analysis for the post-selection protocol}

In the asymptotic security analysis of the CV-QKD system with the post-selection (or the measurement-based NLA), the computed key rate must be multiplied by the success probability of the post-selection, $P_s$, of Eq.~(\ref{succ-prob}). Explicitly, the asymptotic key rate of the post-selection protocol which is secure against Gaussian collective attacks in the reverse reconciliation scenario is given by $K_{\rm ps} = P_s [\beta I_{\rm ps}(a{:}b)-\chi_{\rm ps}(b{:}E)]$, where $I_{\rm ps}(a{:}b)$ is the classical mutual information between Alice and Bob following the post-selection, and $\chi_{\rm ps}(b{:}E)$ is the Holevo bound, i.e., the upper bound on Eve's information on the post-selected data (see Appendix~\ref{AppendixA} and Appendix~\ref{AppendixB} for the key-rate calculation). % calculated using the same method as discussed in Sec.~\ref{Asymptotic} with $V$, $T$ and $\xi$ being replaced by $V_g$, $T_g$ and $\xi_g$, respectively.

In the finite-size security analysis of the CV-QKD protocol with the post-selection, the size of the data contributing to the secret key is no longer $N$. In fact, only the accepted data of size $N_{\rm ps} = P_s N$ contributes to the final post-selected key rate, hence, in order to compute the finite-size key length, the number $N$ must be replaced by $N_{\rm ps} $. Explicitly, the finite-size key length of the post-selection protocol which is secure against Gaussian collective attacks in the reverse reconciliation scenario is given by
\begin{equation}\label{key-length-NLA}
\begin{array}{l}
 \ell_{\rm ps}  \le N_{\rm ps}[ \beta I_{\rm ps}(a{:}b){-}\chi_{\rm ps}(b{:}E) ] {-} \sqrt{ N_{\rm ps}} {\Delta _{\rm ps}} {-} 2\log_2 (\frac{1}{{2\bar \epsilon }}),
 \end{array}
\end{equation}
where ${\Delta _{\rm ps}}$ is calculated using Eq.~(\ref{delta-AEP}) with $N$ being replaced by $N_{\rm ps}$. Hence, the finite-size key rate of the post-selection protocol is given by $K^{\rm FS}_{\rm ps} = \ell_{\rm ps}/N$ or
\begin{equation}\label{key-rate-NLA}
\begin{array}{l}
 K^{\rm FS}_{\rm ps}  {\le} P_s[ \beta I_{\rm ps}(a{:}b) {-} \chi_{\rm ps}(b{:}E) ] {-} \sqrt {\frac{P_s}{N}} {\Delta _{\rm ps}} {-} \frac{2}{N}\log_2 (\frac{1}{{2\bar \epsilon }}).
 \end{array}
\end{equation}
Note that in contrast to the asymptotic regime, the success probability of the post-selection does not affect the finite-size key rate as only a proportional factor. Note also that in Eq.~(\ref{key-length-NLA}) we have again assumed almost the whole raw key of size $N_{\rm ps} $ after the post-selection can be used for secret key generation. However, if the data points of size $k$ are disclosed after the post-selection for the parameter estimation, a classical data of size $N'_{\rm ps}=N_{\rm ps}-k$ is used for the key extraction. In this case, the finite-size key rate is given by $\ell_{\rm ps}/N$, where $\ell_{\rm ps}$ is given by Eq.~(\ref{key-length-NLA}), but now $N_{\rm ps}$ in Eq.~(\ref{key-length-NLA}) has to be replaced by $N'_{\rm ps}$.

%In fact, increasing the success probability always improves the asymptotic key rate, however,

%Although in the asymptotic regime the success probability is only a multiplication coefficient and cannot transform a negative key rate into a positive one.

\section{Numerical results}

\subsection{Gaussian post-selection}

%$\gamma_c \ge 3 g \sqrt{V_B}$ (with $V_B$ the quadrature variance of Bob's measurement outcomes before the post-selection) which means the cut-off circle can embrace at least 99.7$\%$ of the amplified distribution

In the post-selection scheme, when the cut-off $\gamma_c$ is chosen sufficiently large such that the cut-off circle can embrace the amplified distribution, we can assume the distribution of the post-selected data remains statistically Gaussian, %, allowing us to only consider the second moments of the post-selected distribution to characterize the covariance matrix between Alice and Bob.
and the post-selection approximates an ideal NLA (which probabilistically converts a Gaussian state into another Gaussian state) \cite{Fiurasek-PRA-2012, Chrzanowski-NaturePhotonics-2014}. Therefore, in the CV-QKD protocol with the Gaussian post-selection, the security can be analysed based on the equivalent scheme, where the classical filter is replaced with a quantum filter (or NLA) before Bob's heterodyne detection (as it has been analysed in \cite{Fiurasek-PRA-2012}), and the covariance matrix of the amplified state shared between Alice and Bob in the equivalent entanglement-based scheme can be calculated using the covariance matrix of the equivalent system with the effective parameters $\chi_g, T_g,\xi_g$ without the post-selection. Note that the covariance matrix calculated based on the effective parameters $\chi_g, T_g,\xi_g$ of Eq.~(\ref{eff-parameters}) is the same as the covariance matrix ${\bf{M}}_{\rm ps}$ of Eq.~(\ref{CM-postselected}) when the cut-off $\gamma_c$ is chosen sufficiently large. % i.e., $\gamma_c \ge 3 g \sqrt{V_B}$.

%Considering this equivalent system we denote the covariance matrix of the entangled states $\rho_{AB}$, $\rho_{AB_1}$, and $\rho_{AB_2}$ (described in Sec.~\ref{System model}) by $M_{AB}^g$, $M_{AB_1}^g$, and $M_{AB_2}^g$, %Since in the equivalent entanglement-based scheme the Gaussian post-selection changes the parameters of $\lambda$, $T$, and $\xi$ to the effective parameters of $\lambda_g$, $T_g$, and $\xi_g$ respectively, where they are still in the form of $M_{AB}$ in Eq.~(\ref{AB-CM}), $M_{AB_1}$ in Eq.~(\ref{AB1-CM}), and $M_{AB_2}$ in Eq.~(\ref{AB2-CM}) respectively, however, $T$ and $\xi$ must be replaced by $T_g$ and $\xi_g$, and $V$ must be replaced by $V_g=(1+\chi_g^2)/(1-\chi_g^2)$.

%\subsection{Numerical Results for the Gaussian Post-selection}

\begin{figure}[t!]
    \begin{center}
      {\includegraphics [width=3.6 in] {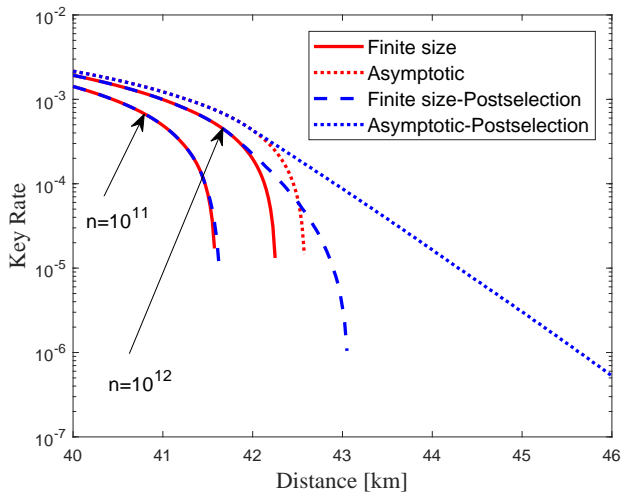}}
    \caption{The achievable secret key rate from reverse reconciliation as a function of channel distance (km) in the no-switching CV-QKD protocol over a lossy channel with $\xi=0.1$ and with 0.2 dB losses per km, without the post-selection (red lines) and with the Gaussian post-selection, where $\gamma_c= 3 g \sqrt{V_B}$ (blue lines) for the asymptotic and finite-size regime ($n=10^{11}$ and $n=10^{12}$) with the discretization parameter of $d=5$ and $\beta=0.95$. The modulation variance (or the squeezing parameter $\chi$) and the gain $g$ are optimised to maximise the key rate.}\label{PS-n}
    \end{center}
\end{figure}

Now we numerically simulate the effect of the Gaussian post-selection on the performance of the CV-QKD protocol in the finite-size regime. In this work, we always consider a lossy quantum channel with 0.2 dB losses per kilometre, and the security parameter $\epsilon=10^{-6}$. We consider different values of block size, $n=10^{11}$ and $n=10^{12}$. Note that the modulation variance (or the squeezing parameter $\chi$ in the equivalent entanglement-based scheme ) and the gain $g$ are optimised to maximise the key rate. We also choose a sufficiently large cut-off \footnote{In our numerical simulations we found that by considering $\gamma_c \ge 3 g \sqrt{V_B}$ \cite{Zhao-PRA-2017}, the covariance matrix of the amplified state, ${\bf{M}}_{\rm ps}$, calculated from Eq.~(\ref{CM-postselected}) is the same as the covariance matrix calculated based on the effective parameters $\chi_g, T_g,\xi_g$ of Eq.~(\ref{eff-parameters}).}, $\gamma_c= 3 g \sqrt{V_B}$  (with $V_B$ the quadrature variance of Bob's measurement outcome before the detection and post-selection) to be able to assume the post-selected state remains Gaussian.
%We indicate $\gamma_c$ the smallest value of cut-off at which the post-selected state is Gaussian.
In Fig.~\ref{PS-n} the achievable secret key rate secure against Gaussian collective attacks is shown as a function of channel distance (km) without the post-selection and with the Gaussian post-selection for both the asymptotic and realistic finite-size regime ($n=10^{11}$ and $n=10^{12}$), and for the realistic reconciliation efficiency of $\beta=0.95$ \cite{experiment-CVQKD-2013}.

We can see from Fig.~\ref{PS-n} that the Gaussian post-selection (blue lines) can be useful when the protocol is operating close to its limit, i.e., in the ``water-fall" region of the key-rate versus distance graph, where modest increases in the correlation between Alice and Bob due to the post-selection (or virtual amplification) can compensate for the sacrificed raw key, allowing the
recovery of a secure key distribution from an initially insecure situation. According to Fig.~\ref{PS-n}, the Gaussian post-selection is able to effectively extend the maximum transmission distance of the CV-QKD protocol in the unrealistic asymptotic regime as it has been previously illustrated in \cite{Fiurasek-PRA-2012, Chrzanowski-NaturePhotonics-2014}. However, in the finite-size regime when the block size is reduced, the improvement of the maximum transmission distance due to the Gaussian post-selection decreases, because increases in the correlation cannot compensate for the scarified raw key. In fact, in the finite-size regime, the improvement of maximum transmission distance due to the Gaussian post-selection can only appear when the block size is sufficiently large (larger than $n=10^{11}$ for the given parameters of Fig.~\ref{PS-n}), and the amount of such an improvement increases with increasing the block size. Note that in Fig.~\ref{PS-n} we have considered a high-noise channel with $\xi=0.1$. We have also performed a further numerical simulation for a lower-noise channel with $\xi=0.05$ (with the other parameters the same as Fig.~\ref{PS-n}). In this case the maximum transmission distance of the protocol is 137.7 km, which can be improved by the Gaussian post-selection for the block sizes larger than $n=10^{15}$. Since we are more interested in a realistic finite-size regime with the block size in the range of $n=10^{8}$ $-$ $10^{12}$, we will consider a higher-noise channel for the rest of our numerical results.

Note that in Fig.~\ref{PS-n}, we have considered the cut-off as $\gamma_c= 3 g \sqrt{V_B}$, so that $ 99.7 \% $ of the amplified distribution lies within the cut-off circle \cite{Zhao-PRA-2017}, and we can assume the post-selected data has a Gaussian distribution which can emulate an ideal NLA. However, if we choose higher values for the cut-off, the post-selection provides a better estimation of the NLA, at the expense of lower success probability. As a result, a larger block size will be required for the CV-QKD performance to be improved by the Gaussian post-selection.

% and finally disappears for the realistic block sizes. Hence, for a realistic finite-size regime, i.e., $n \sim 10^6 - 10^9$ (where the maximum transmission distance is lower than that of the asymptotic regime), the Gaussian post-selection is no longer useful because increases in the correlation cannot compensate for the scarified raw key. For instance, in Fig.~\ref{PS-n} if we choose a realistic block size of  $n=10^{9}$, the maximum transmission distance is reduced to $11.4$ km which is the same for both the protocols with and without the Gaussian post-selection.

\subsection{Non-Gaussian post-selection}

In the measurement-based NLA the choice of the filter cut-off, $\gamma_c$, is critical. Larger cut-off will improve the approximation of the ideal NLA, however, a cut-off that is too high will unnecessarily sacrifice raw data, and decrease the success probability. On the other hand, a cut-off that is too low will increase the success probability, at the expense of reducing the mutual information between Alice and Bob. According to our numerical results for the Gaussian post-selection, the success probability plays a significant role in the finite-size security analysis, since the success probability determines the size of data which contributes to the post-selected key. In this section we investigate whether a reduction of the post-selection cut-off (which will increase the success probability) improves the post-selection performance in the finite-size regime.

% Let us consider a quantum channel equivalent with an optical fiber of 45 km, and the excess noise of $\xi=0.1$. For this channel the optimized Gaussian post-selection cannot generate positive finite key rates with a block size of $n=10^{14}$ (or smaller block size of $n<10^{14}$, see Fig.~\ref{PS-n}).  For this quantum channel we now investigate the effect of a reduction of the post-selection cut-off on the CV-QKD performance. When the cut-off decreases to the lower values of $\gamma^G_c$, the post-selected state is not Gaussian anymore. However, it has been proved that for all bipartite quantum states $\rho_{AB}$ with covariance matrix $M_{AB}$, one has $K(\rho_{AB}) \ge K(\rho^G_{AB})$, where $\rho^G_{AB}$ is the Gaussian state with the same $M_{AB}$. This means that $K(\rho^G_{AB})$ is a \emph{lower} bound on the secret key rate for any protocol (even non-Gaussian) and collective attack (including non-Gaussian). Thus, by calculating the covariance matrix of the post-selected state using Eq.~(\ref{CM-postselected}) we can compute a lower bound on the post-selected key rate in the non-Gaussian regime.

When the fiter cut-off decreases from $\gamma_c = 3 g \sqrt{V_B}$, the statistics of the post-selected data start changing from Gaussian to non-Gaussian. However, based on the optimality of Gaussian attacks \cite{Gaussian-optimality-1, Gaussian-optimality-2, Gaussian-optimality-3}, for all bipartite quantum states $\rho_{AB}$ with covariance matrix ${\bf{M}}_{AB}$, one can maximise Eve's information by considering $\rho^G_{AB}$, which is the Gaussian state having the same covariance matrix ${\bf{M}}_{AB}$.
%has $K(\rho_{AB}) \ge K(\rho^G_{AB})$, where $\rho^G_{AB}$ is the Gaussian state having the same covariance matrix ${\bf{M}}_{AB}$. This means that $K(\rho^G_{AB})$ is a lower bound on the secret key rate for any protocol (even non-Gaussian) and collective attack (including non-Gaussian).
Hence, in order to analyse the security of the protocol in the non-Gaussian regime, we only require to calculate the covariance matrix of the non-Gaussian amplified state. Note that when the post-selection is in the non-Gaussian regime, we cannot use Eq.~(\ref{eff-parameters}) anymore to calculate the covariance matrix of the amplified state. Instead, we have to use the Q-function of the post-selected state, i.e., Eq.~(\ref{CM-postselected}) to calculate the covariance matrix of the amplified state, and compute a lower bound on the post-selected key rate. Note also that the technique of the measurement-based NLA with an entangled-state input has always been investigated in the Gaussian regime, where the filter cut-off is sufficiently large \cite{Fiurasek-PRA-2012, Chrzanowski-NaturePhotonics-2014}. However, here we investigate the characterization of the measurement-based NLA with an entangled-state input in the non-Gaussian regime by decreasing the filter cut-off, and the impact this cut-off reduction can have on the related CV-QKD performance.

Let us consider a quantum channel equivalent with an optical fiber of 43 km, which according to Fig.~\ref{PS-n}, is almost the maximum transmission distance of the CV-QKD system with the optimised Gaussian post-selection, where we have the excess noise of $\xi=0.1$, $\beta=0.95$, and the block size of $n=10^{12}$. For this channel the optimized Gaussian post-selection (with $\gamma_c = 3 g \sqrt{V_B}$) generates the finite-size key rate of $K^{\rm FS}_{\rm ps}=3.4 \times 10^{-6}$ (bits per symbol).  For this quantum channel we now investigate the effects a decrease in the post-selection cut-off, $\gamma_c$, can have on the CV-QKD performance.

\begin{figure}[t!]
    \begin{center}
      {\includegraphics [width=3.6 in] {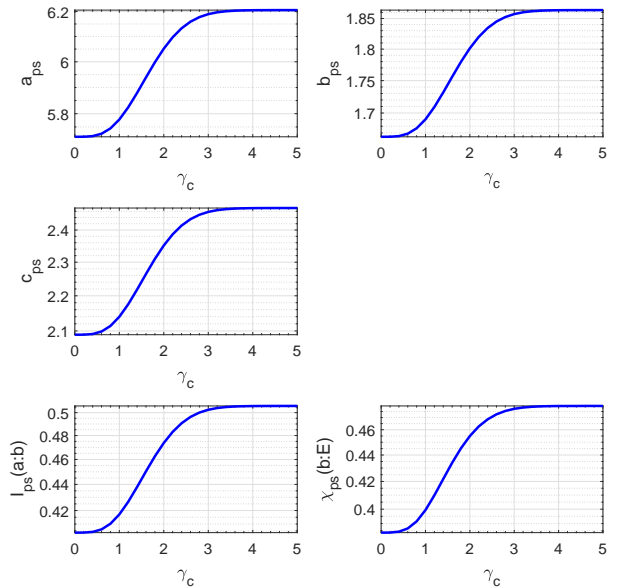}}
    \caption{The elements of the covariance matrix, ${\bf{M}}_{\rm ps}$, of the post-selected state, (i.e., $a_{\rm ps}$, $b_{\rm ps}$, and $c_{\rm ps}$), the classical mutual information between Alice and Bob, $I_{\rm ps}(a{:}b)$, and Eve's information from reverse reconciliation (i.e., the Holevo bound), $\chi_{\rm ps}(b{:}E)$ as a function of the filter cutoff $\gamma_c$ for a quantum channel equivalent with an optical fiber of 43 km, $\xi=0.1$, $\beta=0.95$, $\chi = 0.8379$, $g=1.1$ and the block size of $n=10^{12}$.}\label{CM_Inf_cutoff}
    \end{center}
\end{figure}

\begin{figure}[t!]
    \begin{center}
      {\includegraphics [width=3.6 in] {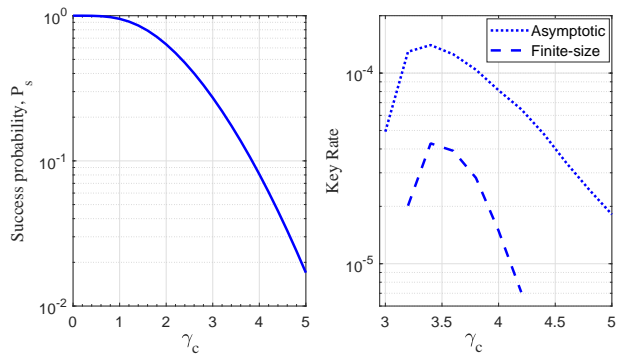}}
    \caption{The post-selection success probability, $P_s$, and the finite-size and asymptotic key rate from reverse reconciliation as a function of the filter cutoff $\gamma_c$ for a quantum channel and the protocol with the same parameters as Fig.~\ref{CM_Inf_cutoff}.}\label{key_cutoff}
    \end{center}
\end{figure}

In Fig.~\ref{CM_Inf_cutoff}, the three top plots show the elements of the covariance matrix, ${\bf{M}}_{\rm ps}$, of the amplified state (i.e., $a_{\rm ps}$, $b_{\rm ps}$, and $c_{\rm ps}$ in Eq.~(\ref{CM-postselected})) as a function of the post-selection cut-off $\gamma_c$. As it can be seen, for $\gamma_c \ge 3 g \sqrt{V_B} = 4.26$, the post-selected state can be assumed to be Gaussian, as the elements of the covariance matrix ${\bf{M}}_{\rm ps}$ remain almost constant and equal to the covariance matrix elements of the amplified state resulted from an ideal NLA (calculated based on Eq.~(\ref{eff-parameters})). We can see the covariance matrix elements of the amplified state  decrease as the cut-off is reduced. As a result, the classical mutual information between Alice and Bob, $I_{\rm ps}(a{:}b)$, as well as Eve's information, i.e., the Holevo bound, $\chi_{\rm ps}(b{:}E)$ (with both being calculated based on the covariance matrix ${\bf{M}}_{\rm ps}$) decrease with the cut-off reducing (shown in the two bottom plots of Fig.~\ref{CM_Inf_cutoff}).
Although the raw key-rate term, $\beta I_{\rm ps}(a{:}b){-}\chi_{\rm ps}(b{:}E)$, also drops with the decrease in the cut-off, the success probability of the post-selection, $P_s$, exponentially increases with the cut-off decreasing according to the left plot of Fig.~\ref{key_cutoff}. As a result, both the asymptotic and finite-size key rates first increase  with the cut-off decreasing up to an optimized value, and then decrease until they disappear (see Fig.~\ref{key_cutoff}, right plot). Therefore, our results show that there is an optimal value for the cut-off in the non-Gaussian regime which maximizes the key rate. According to Fig.~\ref{key_cutoff}, the finite-size key rate can be improved up to $K^{\rm FS}_{\rm ps}=4.3 \times 10^{-5}$ (i.e., an improvement of more than one order of magnitude) by decreasing the cut-off from Gaussian regime to non-Gaussian regime, i.e., from $\gamma_c = 3 g \sqrt{V_B} = 4.26$ to $\gamma_c = 3.4$.  Note that in Figs.~\ref{CM_Inf_cutoff} and \ref{key_cutoff}, for $\gamma_c \ge 3 g \sqrt{V_B}$, the post-selected state remains almost Gaussian and the post-selection can emulate the ideal NLA, while $\gamma_c=0$ corresponding to no post-selection.

Note that the lower bound on the key rate which we calculate for the post-selection in the non-Gaussian regime is not tight, and it could likely be improved using the numerical approach of \cite{Norbert2019}. The bound is loose because it relies on Gaussian optimality proof \cite{Gaussian-optimality-1, Gaussian-optimality-2, Gaussian-optimality-3}, which means that $\chi_{\rm ps}(b{:}E)$ is computed for the Gaussian state with the same covariance matrix as the non-Gaussian amplified state, and $\chi_{\rm ps}(b{:}E)$ is therefore overestimated. Techniques that provide tighter bounds for non-Gaussian statistics would therefore result in a smaller value for the optimal cut-off which, given the exponential improvement in the fraction of data kept, could significantly improve the key rates.

\begin{figure}[t!]
    \begin{center}
      {\includegraphics [width=3.6 in] {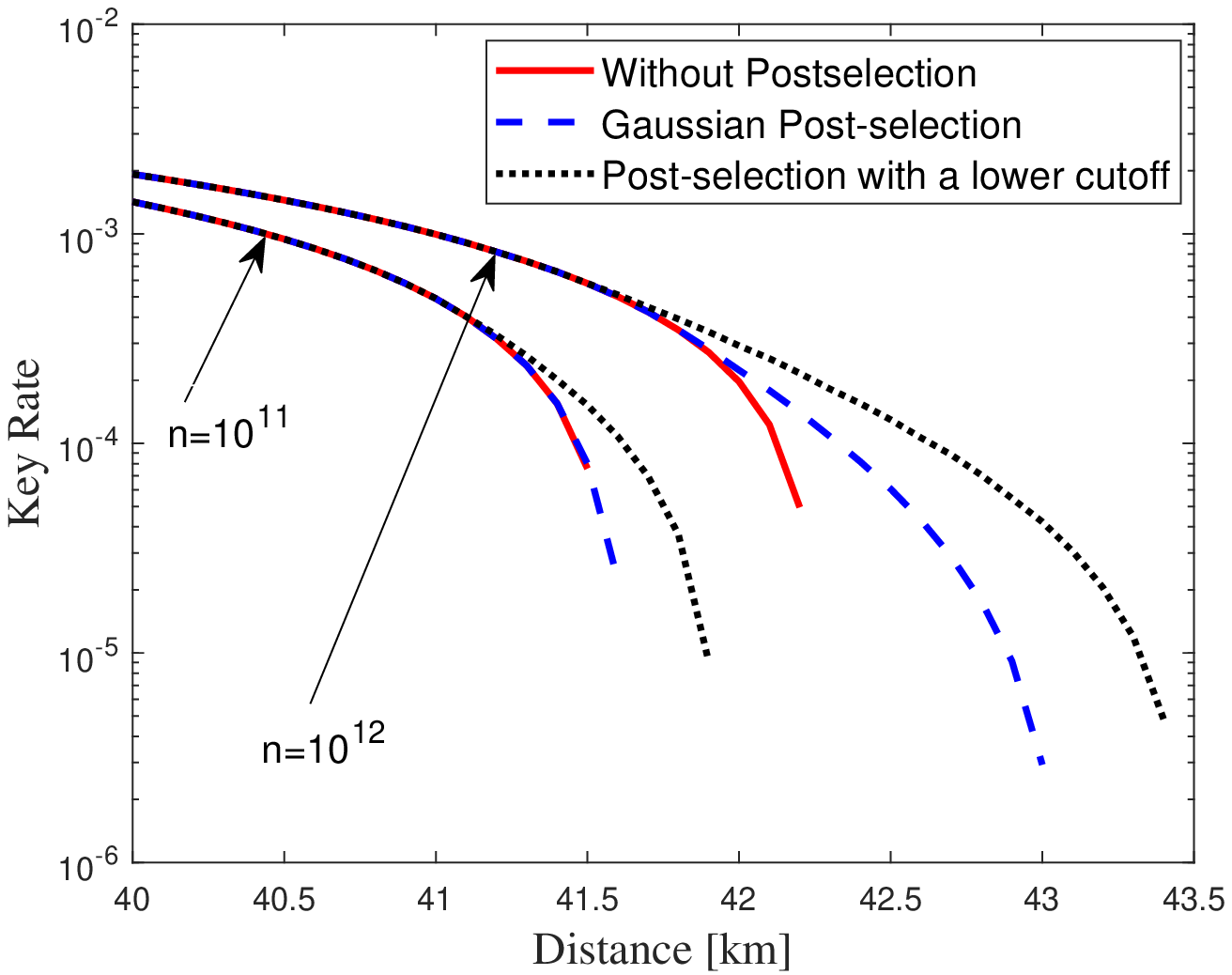}}
    \caption{The achievable secret key rate from reverse reconciliation as a function of channel distance (km) in the CV-QKD protocol over a lossy channel with $\xi=0.1$ and with 0.2 dB losses per km, without post-selection (red lines), with the Gaussian post-selection, i.e., $\gamma_c= 3 g \sqrt{V_B}$ (blue lines), and with the non-Gaussian post-selection, i.e., choosing a lower cut-off $\gamma_c= 2.5 g \sqrt{V_B}$ (black lines) for the finite-size regime ($n=10^{11}$ and $n=10^{12}$) with $\beta=0.95$.}\label{PS-n-decrease}
    \end{center}
\end{figure}

Now we repeat our numerical simulations for the post-selection in Fig.~\ref{PS-n}, with a lower cut-off, $\gamma_c= 2.5 g \sqrt{V_B}$, and compute the post-selected finite-size key rate, with the results shown in Fig.~\ref{PS-n-decrease}. We can see that decreasing the cut-off from $\gamma_c= 3 g \sqrt{V_B}$ to $\gamma_c= 2.5 g \sqrt{V_B}$ has a positive impact on the CV-QKD performance, including the improvement of the finite-size key rate by up to an order of magnitude at the maximum transmission distance of the protocol, as well as the extension of the transmission distance up to a half kilometre. We can see that for $n=10^{12}$ by decreasing the cut-off from $\gamma_c= 3 g \sqrt{V_B}$ to $\gamma_c= 2.5 g \sqrt{V_B}$ the improvement of the transmission distance due to the post-selection becomes more significant. Furthermore, we can see that while for $n=10^{11}$ there is no improvement in the transmission distance due to the post-selection with $\gamma_c= 3 g \sqrt{V_B}$, the transmission distance can be improved by decreasing the cut-off to $\gamma_c= 2.5 g \sqrt{V_B}$. Recall again that here in our numerical simulations for $\gamma_c= 2.5 g \sqrt{V_B}$ the post-selected state is not Gaussian (although it is close to the Gaussian regime), hence we use the output $Q$-function, $Q_{\rm out(\alpha,\beta)}$, to calculate the elements of the covariance matrix of the post-selected state, ${\bf{M}}_{\rm ps}$. Our results show the importance of the proper choice of the post-selection cut-off in the CV-QKD system.

% Note that in our numerical simulations, the success probability in the worst case of large $\gamma_c=5$ is almost $P_s = 10 ^ {-2}$. Considering our large block sizes of $n=10 ^ {11}$, or $n=10 ^ {12}$, after the post-selection we will have the raw post-selected key of size $N_{\rm ps}=10 ^ {9}$ or $N_{\rm ps}=10 ^ {10}$. If we can use either almost the whole post-selected data  or even half post-selected data for the parameter estimation (and the secret key extraction), the block size used for the parameter estimation would be still large and we can assume the estimated covariance matrix is the same as the expected covariance matrix with a good precision.

Additional calculations beyond those illustrated here have been carried out covering direct reconciliation, which results in similar trends to those indicated here. However, direct reconciliation is only successful when the channel loss is below 3 dB. In the direct reconciliation, Eve's information should be calculated based on the mutual information between Alice and Eve, i.e., $\chi_{\rm ps}(a{:}E)$. For the numerical simulations of the direct reconciliation see Appendix~~\ref{AppendixC}.

\subsection{Parameter estimation in the post-selection protocol}

Note that the no-switching CV-QKD protocol is experimentally implemented in the prepare-and-measure (PM) scheme, where for the post-selection the classical filter is applied on Bob's heterodyne detection results, while for the security analysis we need to know the covariance matrix, ${\bf{M}}_{\rm ps}$, of the amplified (or post-selected) state shared between Alice and Bob in the equivalent entanglement-based (EB) scheme.

In the case of Gaussian post-selection, we can consider a normal linear model for Alice and Bob's post-selected variables in the PM scheme, $x^{\rm PM}_{\rm ps}$ and $y^{\rm PM}_{\rm ps}$, respectively  as $y^{\rm PM}_{\rm ps} = t_g  x^{\rm PM}_{\rm ps} + z_{\rm ps}$, where $t_g = \sqrt {\frac{T_g}{2}}$, and $z_{\rm ps}$ follows a centred normal distribution with unknown variance $\sigma_g^2 = 1 + \frac{1}{2} {T_g \xi_g} $ (note that Alice's variable $x^{\rm PM}_{\rm ps}$ has the variance $V^g_A$). The maximum-likelihood estimators for the effective parameters, $t_g$, $\sigma_g^2$, and $V^g_A$ are given by \cite{MLE-estimator2010,MLE-estimator2012}
\begin{equation}\label{b}
\begin{array}{l}
 \hat t_g = \frac{{\sum\nolimits_{i = 1}^{k} {{{x_{i}}}{{y_i}}} }}{{\sum\nolimits_{i = 1}^{k} {{{{{x^2_i}}}}} }}, \\
 \\
 {\hat \sigma_g^2} = \frac{1}{{k}}\sum\nolimits_{i = 1}^{k} {{{({{y_{i}}} - \hat t_g{{x_{i}}})}^2}},\\
 \\
\hat V^g_A = \frac{1}{{k}}\sum\nolimits_{i = 1}^{k} {x_i^2},
 \end{array}
\end{equation}
with the uncertainty in the effective parameters expressed as \cite{MLE-estimator2010,MLE-estimator2012}
\begin{equation}\label{b-ci}
\begin{array}{l}
\Delta (t_g) = {z_{\epsilon_{\rm PE} /2}}\sqrt {\frac{{{\hat \sigma_g ^2}}}{{\sum\nolimits_{i = 1}^{k} {{{{{x_i^2}}}}} }}},  \\
\\
\Delta ({{\sigma }_g^2}) = {z_{\epsilon_{\rm PE} /2}}\frac{{{\hat \sigma_g ^2}\sqrt 2 }}{{\sqrt {k} }},\\
\\
\Delta (V^g_A) = {z_{\epsilon_{\rm PE} /2}}\frac{{{\hat V_A^g }\sqrt 2 }}{{\sqrt {k} }}
 \end{array}
\end{equation}
where $x_i$ and $y_i$ are the realizations of  $x^{\rm PM}_{\rm ps}$ and $y^{\rm PM}_{\rm ps}$, respectively, and $k$ is the number of data points randomly chosen from the post-selected data for the parameter estimation \footnote{Note that ${z_{\epsilon_{\rm PE} /2}}$ is such that $1-{\rm erf}(\frac{{z_{\epsilon_{\rm PE} /2}}}{\sqrt{2}})/2 = \epsilon_{\rm PE} /2$.}.
As a result, the covariance matrix , ${\bf{M}}_{\rm ps}$, of the amplified state shared between Alice and Bob in the EB scheme, which maximises Eve's information \cite{MLE-estimator2010} is given by ${\bf{\hat M}}_{\rm ps} = [\hat a_{\rm ps}{\bf{I}}_2,\hat c_{\rm ps}{\bf{Z}};\hat c_{\rm ps}{\bf{Z}},\hat b_{\rm ps}{\bf{I}}_2]$, where
\begin{equation}\label{AB1-CM-subchannel-estimation}
\begin{array}{l}
\hat a_{\rm ps} = V^g_{A,\rm{max}} {+}1,\\
\\
\hat b_{\rm ps} = 2({t_{g,\rm{min}}^2 V^g_{A,\rm{max}} {+} \sigma_{g,\rm{max}}^2}){-}1 ,\\
\\
\hat c_{\rm ps} = \sqrt{2} \, t_{g,\rm{min}} \sqrt {{V^g_{A,\rm{max}}}^2 {+} 2V^g_{A,\rm{max}} },
\end{array}
\end{equation}
and where
\begin{equation}\label{estimation-subchannel}
\begin{array}{l}
 t_{g,\rm{min}} = \hat t_{g} - \Delta (t_{g}) \\
\\
\sigma_{g,\rm{max}}^2 = \hat \sigma_{g}^2 + \Delta ({{\sigma }_{g}^2}),\\
\\
V^g_{A,\rm{max}} = \hat V^g_{A} + \Delta(V^g_{A}).
 \end{array}
\end{equation}

However, in the case of non-Gaussian post-selection, the relation between the cross-correlation term, $c_{\rm ps}$, of the covariance matrix ${\bf{M}}_{\rm ps}$ in the EB scheme is not directly related to the cross-correlation term of the data observed by Alice and Bob in the PM scheme, i.e., $\frac{1}{k}{\sum\nolimits_{i = 1}^{k} {{{x_{i}}}{{y_i}}} }$.
%Thus, in the PM scheme Alice and Bob need to perform parameter estimation before post-selection over the whole data to estimate channel parameters to make sure the estimated channel parameters are the same as ones their expectations.
%the covariance matrix, ${\bf{M}}$, of the state shared between Alice and Bob in the equivalent entanglement-based (EB) scheme before post-selection is the same as the one expected from a Gaussian attack.
Hence, instead of calculating ${\bf{M}}_{\rm ps}$ from the data observed in the PM scheme, Alice and Bob can first reconstruct the equivalent data in the EB scheme based on the whole data from the PM scheme preceding the post-selection. Considering Alice and Bob's variables in the PM scheme preceding the post-selection as $x^{\rm PM}$ and $y^{\rm PM}$, Alice and Bob's variables in the equivalent EB scheme preceding the post-selection would be $x^{\rm EB} =\frac{\sqrt{V_A+2}}{\sqrt{2 V_A}} x^{\rm PM}$ and $y^{\rm EB} = y^{\rm PM}$, with $V_A$ is the initial modulation variance in the PM scheme preceding the post-selection. Next, Bob applies the classical filter on his data and publicly reveals whether the data is kept or rejected. Finally, Alice and Bob perform parameter estimation over a randomly-chosen subset (of size $k$) of their post-selected data, $x^{\rm EB}_{\rm ps}$ as $y^{\rm EB}_{\rm ps}$, to directly estimate ${\bf{M}}_{\rm ps}$ via $\frac{1}{{k}}\sum\nolimits_{i = 1}^{k} {{x'_i}^2} $, $\frac{1}{{k}}\sum\nolimits_{i = 1}^{k} {{y'_i}^2}$, and $\frac{1}{{k}}\sum\nolimits_{i = 1}^{k} {{x'_i}{y'_i}}$, where $x'_i$ and $y'_i$ are the realizations of  $x^{\rm EB}_{\rm ps}$ and $y^{\rm EB}_{\rm ps}$, respectively.

\section{Conclusions}
In this work we have investigated the impact post-selection or measurement-based NLA can have on the CV-QKD performance (when it is applied on the measurement outcome of Bob's detection) in the finite-size regime. We found that the post-selection can extend the maximum transmission distance of CV-QKD protocol in the finite-size regime providing the finite block size is sufficiently large.
For finite blocks with a practical length, we found that the post-selection is effective for the protocols with high values of excess noise. Further, we analysed the performance of the measurement-based NLA on the entangled-state input in the non-Gaussian regime by decreasing the post-selection cut-off, thereby illustrating that there is an optimal value for the post-selection cut-off that optimises the CV-QKD performance in terms of both the finite key rate and transmission range.

\section{Acknowledgements}
The authors acknowledge valuable discussions with Austin Lund. This research was supported by funding from the Australian Department of Defence. This research is also supported by the Australian Research Council (ARC) under the Centre of Excellence for Quantum Computation and Communication Technology (Project No. CE170100012). NW acknowledges funding support from the European Unions Horizon 2020 research and innovation programme under the Marie Sklodowska-Curie grant agreement No.750905 and Q.Link.X from the BMBF in Germany.

%\newpage
\appendix

\section{Calculation of mutual information and Holevo bound}\label{AppendixA}

In the entanglement-based scheme of the no-switching CV-QKD protocol, Alice generates a pure two-mode Gaussian entangled state, i.e., a two-mode squeezed vacuum state with the quadrature variance $V$. % is completely described by its first moment, which is zero, and its covariance matrix,
%\begin{eqnarray}\label{AB-CM}
%{\bf{M}}_{AB}= \left[ {\begin{array}{*{20}{c}}
%{V\,\bf{I}}&{\sqrt {{V^2} - 1} \,\bf{Z}}\\
%{\sqrt {{V^2} - 1} \,\bf{Z}}&{V\,\bf{I}}
%\end{array}} \right] .
%\end{eqnarray}
Alice keeps the first mode and transmits the second mode through a quantum channel with transmissivity $T$ and excess noise $\xi$. The covariance matrix of the mixed state ${\rho _{A{B}}}$ at the output of the channel before the detection is given by
\begin{equation}\label{AB1-CM}
{{\bf{M}}} = \left[ {\begin{array}{*{20}{c}}
{V\,{\bf{I}}_2}&{\sqrt {T \,} \sqrt {{V^2} - 1}\,\bf{Z}}\\
{\sqrt T  \,\sqrt {{V^2} - 1}\,\bf{Z}}&{\left( {T (V + {\chi _{\rm line}})} \right)\,{\bf{I}}_2}
\end{array}} \right],
\end{equation}
where ${\chi _{\rm line}} = \xi  + \frac{{1}}{T}-1$. Having the covariance matrix ${\bf{M}}$, we are able to compute the Q-function, $Q_{\rm in}(\alpha_m,\beta_m)$, of the state shared between Alice and Bob preceding the post-selection using Eq.~(\ref{input-Qfunc}). Then, following Eq.~(\ref{succ-prob}) and Eq.~(\ref{output-Qfunc}) we can compute the post-selection success probability $P_s$ as well as the Q-function, $Q_{\rm out}(\alpha,\beta)$, of the post-selected state, from which we can compute the inferred covariance matrix, ${{\bf{M_{\rm ps}}}}$, of the amplified state using Eq.~(\ref{CM-postselected}).

Following the post-selection, the mutual information between Alice and Bob, $I_{\rm ps}(a{:}b)$, can be calculated as (see Appendix~\ref{AppendixB} for the actual mutual information)
\begin{equation}\label{mutual-G}
I_{\rm ps}(a{:}b) = {\log _2}\frac{ a_{\rm ps} + 1 }{ a_{\rm ps} + 1 - \frac{c^2_{\rm ps}}{b_{\rm ps}+1}}.
\end{equation}

%$I_{\rm ps}(a{:}b) = {\log _2}\frac{ V+{\chi _{\rm line}} + \frac{1}{T }}{ 1+{\chi _{\rm line}} + \frac{1}{T }}$.

%, where $V_{B_2^{\rm het}}$ is the variance of heterodyne-detected mode $B_2$, and is given by $V_{B_2^{\rm het}}= T (V+\chi_{\rm tot})/2$, where $\chi_{\rm tot}  = {\chi _{\rm line}} + \frac{1}{T }$. The conditional variance $V_{{B_2}^{\rm het}|A^{\rm het}}$ is the variance of heterodyne-detected mode $B_2$ conditioned on Alice's heterodyne detection of mode $A$, which is given by $V_{{B_2}^{\rm het}|A^{\rm het}} =  T (1+{\chi _{\rm line}} + \frac{1}{T })/2$.

%In the individual attack, the Shannon mutual information $I(B:E)$ is given by $I(B:E) = {\log _2}\frac{{{V_{B_2^{\rm het}}}}}{{{V_{{B_2}^{\rm het}\left| E \right.}}}}$, where $V_{{B_2^{\rm het}}|E}$ in the case of Bob's detection noise not being accessible to Eve\footnote{Note that if Bob's detection noise is accessible to Eve, we have $I(B:E) = {\log _2}\frac{{{V_{B_2^{\rm het}}}}}{{{V_{{B_1}^{\rm het}\left| E \right.}}}}$, where $V_{{B_1}^{\rm het}|E}=[\frac{V{x_E}+1}{V+x_E}+1]/2$ \cite{thesis,inefficient_heterodyne, individula-2007-1, individula-2007-2}.} is given by $V_{{B_2}^{\rm het}|E}=\eta[\frac{V{x_E}+1}{V+x_E}+\chi_{\rm het}]/2$, where $x_E=T(2-\xi)^2/(\sqrt{2-2T+T\xi}+\sqrt{\xi}))^2+1$ \cite{inefficient_heterodyne, thesis, individula-2007-1, individula-2007-2}.%, and $\chi_{\rm het}=[1 + (1 - \eta ) + {\upsilon _{el}}]/\eta$.

In the collective attack, the Holevo mutual information $\chi(b{:}E)$ is given by $\chi(b{:}E)=S(\rho_E)-S(\rho_{E|b})$, where $S(\rho )$ is the von Neumann entropy of the state $\rho$.
%Here we assume Bob's detection noise is not accessible to Eve.
Note that $S(\rho_E)$ is given by the von Neumann entropy of the amplified state, which can be calculated through the symplectic eigenvalues $\lambda _{1,2}$ of covariance matrix ${{\bf{M_{\rm ps}}}}$\footnote{The von Neumann entropy of an $n$-mode Gaussian state $\rho$ with the covariance matrix $\bf{M}$ is given by $S(\rho)=\sum\nolimits_{i = 1}^n {G(\frac{\lambda _i-1}{2})} $, where $\lambda _{i}$ are the symplectic eigenvalues of the covariance matrix ${\bf{M}}$, and $G(x) = (x + 1){\log _2}(x + 1) - x {\log _2}(x)$.}. The second entropy $S(\rho_{E|b})$ is given by the von Neumann entropy of Alice's state conditioned on Bob's detection, which can be calculated through the symplectic eigenvalue of the covariance matrix of the conditional state  ${\bf{M}}_{A|b}= {\bf{A_{\rm ps}}}-{\bf{C_{\rm ps}}} \,\, ({\bf{B_{\rm ps}}}+{\bf{I}}_2)^{-1} \,\, {\bf{C}}^T_{\rm ps}$, where ${\bf{A_{\rm ps}}} = a_{\rm{ps}}{\bf{I}}_2$, ${\bf{B_{\rm ps}}} = b_{\rm{ps}}{\bf{I}}_2$, and ${\bf{C_{\rm ps}}} = c_{\rm{ps}}{\bf{Z}}$.

%where ${\bf{H}}_{\rm het} = ({\bf{M}}_{B_2}+\bf{I})^{-1}$, and where ${\bf{M}}_{B_2} = {\left( {\eta T (V + \chi_{\rm t} )} \right)\,\bf{I}}$. Note that the matrices ${\bf{M}}_{AFG}, \boldsymbol{\sigma}_{AFG,{B_2}}$, and ${\bf{M}}_{B_2}$ can be derived from the decomposition of the covariance matrix\footnote{Note that if Bob's detection noise is accessible to Eve, $S(\rho_E) = S(\rho_{AB_2})$, and $S(\rho_{E|B}) = S(\rho_{E|B_2}) = S(\rho_{A|B_2})$.}
%\begin{equation}\label{big-CM}
%{{\bf{M}}_{AFG{B_2}}} = \left[ {\begin{array}{*{20}{c}}
%   {{{\bf{M}}_{AFG}}} & {\boldsymbol{\sigma} _{AFG,{B_2}}}  \\
%   {{\boldsymbol{\sigma}^T_{AFG,{B_2}}}} & {{{\bf{M}}_{{B_2}}}}  \\
%\end{array}} \right].
%\end{equation}
%Note that the covariance matrix ${\bf{M}}_{AFG{B_2}}$ is given by ${\bf{M}}_{AFG{B_2}} = {\bf{B}}^T [{\bf{M}}_{A{B_1}}  \oplus {\bf{M}}_{{F_0}G} ]{\bf{B}}$, where ${\bf{B}}$ is the beam splitter transformation matrix given by
%\begin{eqnarray}\label{BS-CM}
%\bf{B}= \left[ {\begin{array}{*{20}{c}}
%{\sqrt{\eta}\,\bf{I}}&{\sqrt {1 - \eta} \,\bf{I}}\\
%{-\sqrt {1 - \eta} \,\bf{I}}&{\sqrt{\eta}\,\bf{I}}
%\end{array}} \right] ,
%\end{eqnarray}
%and the covariance matrix of the EPR state $\rho_{{F_0}G}$ is given by
%\begin{eqnarray}\label{F0G-CM}
%{\bf{M}}_{{F_0}G}= \left[ {\begin{array}{*{20}{c}}
%{\upsilon\,\bf{I}}&{\sqrt {{\upsilon^2} - 1} \,\bf{Z}}\\
%{\sqrt {{\upsilon^2} - 1} \,\bf{Z}}&{\upsilon\,\bf{I}}
%\end{array}} \right] .
%\end{eqnarray}

\section{Actual mutual information}\label{AppendixB}

\begin{figure}[t!]
    \begin{center}
      {\includegraphics [width=3.6 in] {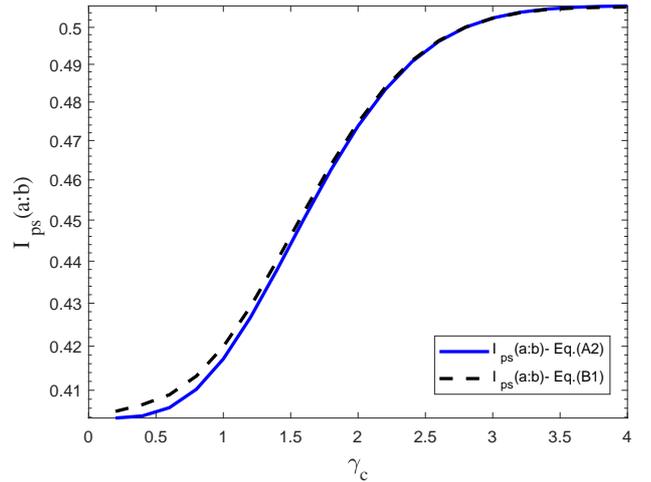}}
    \caption{The classical mutual information between Alice and Bob following the post-selection, $I_{\rm ps}(a{:}b)$, using the covariance matrix, ${{\bf{M_{\rm ps}}}}$, of the amplified state via Eq.~(\ref{mutual-G}) (solid line), and using the Q-function of the post-selected state via Eq.~(\ref{mutual-nG}) (dashed line), as a function of the filter cutoff $\gamma_c$ for a quantum channel equivalent with an optical fiber of 43 km, $\xi=0.1$, $\chi = 0.8379$, $g=1.1$.}\label{mutual-info}
    \end{center}
\end{figure}

\begin{figure}[t!]
    \begin{center}
      {\includegraphics [width=3.6 in] {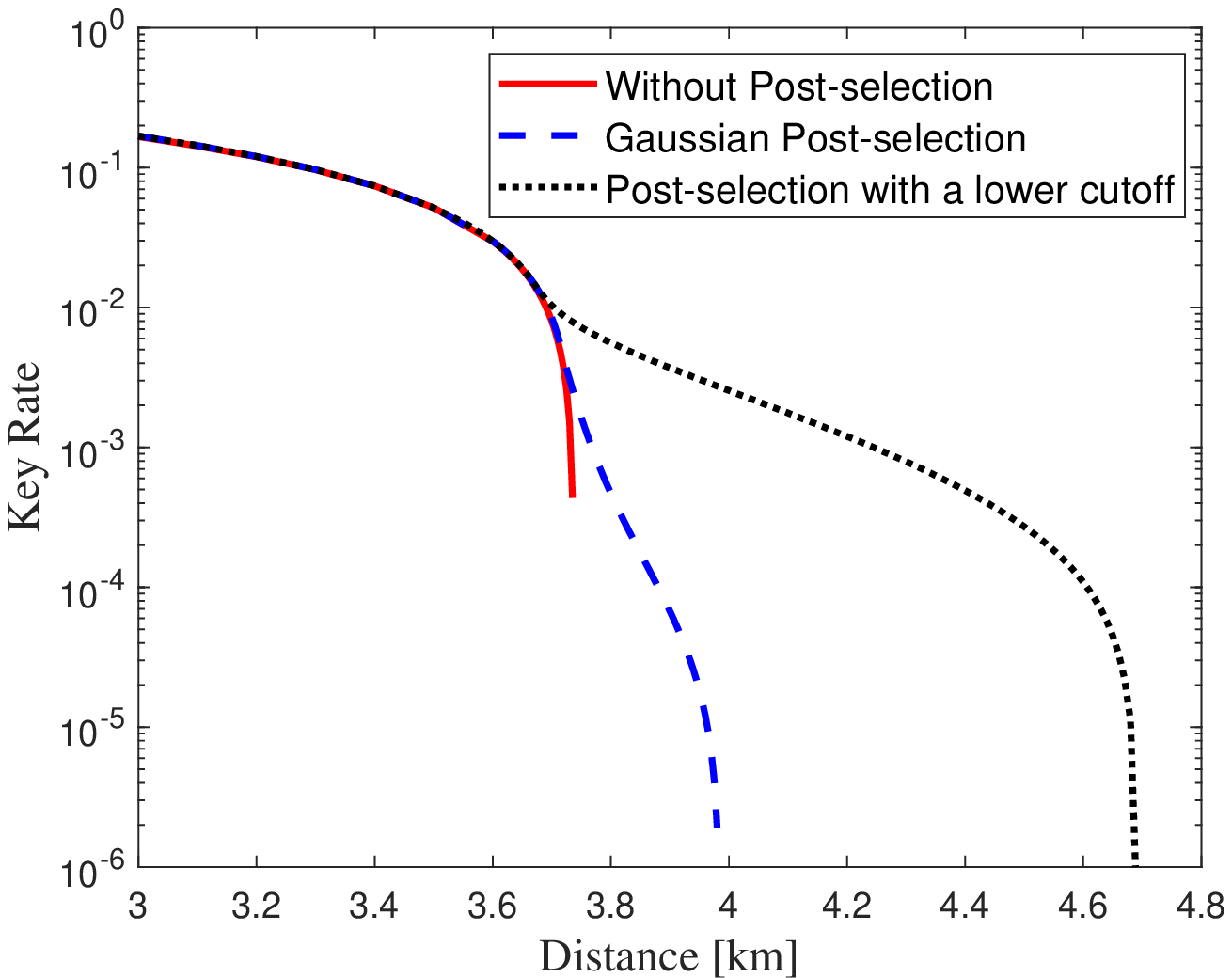}}
    \caption{The achievable secret key rate from direct reconciliation as a function of channel distance (km) in the CV-QKD protocol over a lossy channel with $\xi=0.1$ and with 0.2 dB losses per km, without post-selection (red lines), with the Gaussian post-selection, i.e., $\gamma_c= 3 g \sqrt{V_B}$ (blue lines), and with the non-Gaussian post-selection, i.e., choosing a lower cut-off $\gamma_c= 2 g \sqrt{V_B}$ (black lines) for the finite-size regime ($n=10^{10}$) with $\beta=0.95$.}\label{PS-DR}
    \end{center}
\end{figure}

In the case of Gaussian post-selection, when the post-selected state has Gaussian statistics, the actual mutual information between Alice and Bob can be calculated using the covariance matrix, ${{\bf{M_{\rm ps}}}}$, of the amplified (or post-selected) state via Eq.~(\ref{mutual-G}). However, in the case of non-Gaussian post-selection, when the post-selected state has non-Gaussian statistics, the actual mutual information can be calculated using
\begin{equation}\label{mutual-nG}
I_{\rm ps}(a{:}b) = H_{\rm ps}(a) + H_{\rm ps}(b)- H_{\rm ps}(a,b),
\end{equation}
where $H_{\rm ps}(a)$ is the Shannon entropy of Alice's classical variable (or Alice's heterodyne-measurement result in the entanglement-based scheme) following the post-selection, $H_{\rm ps}(b)$ is the Shannon entropy of Bob's heterodyne-measurement result following the post-selection, and $H_{\rm ps}(a,b)$ is  the joint entropy of Alice and Bob's classical variables following the post-selection. In Eq.~(\ref{mutual-nG}), $H_{\rm ps}(a,b)$ is calculated as
\begin{equation}\label{H(a,b)}
\begin{array}{l}
H_{\rm ps}(a,b) = - {\int}{\int} d^2{\alpha}{\int}{\int} d^2{\beta}\,\,Q_{\rm out}({\alpha},{\beta}) \log_2[Q_{\rm out}({\alpha},{\beta})]
\end{array}
\end{equation}
where $Q_{\rm out}({\alpha},{\beta})$ is the Q-function of the post-selected state given by Eq.~(\ref{output-Qfunc}). In Eq.~(\ref{mutual-nG}), $H_{\rm ps}(b)$ is calculated as
\begin{equation}\label{H(b)}
\begin{array}{l}
H_{\rm ps}(b) = - {\int}{\int} d^2{\beta}\,\,Q_{\rm out}({\beta}) \log_2[Q_{\rm out}({\beta})]
\end{array}
\end{equation}
where $Q_{\rm out}({\beta})$ is the Q-function of Bob's post-selected state, given by
\begin{equation}\label{Q(beta)}
\begin{array}{l}
Q_{\rm out}({\beta}) = {\int}{\int} d^2{\alpha}\,\,Q_{\rm out}({\alpha},{\beta}).
\end{array}
\end{equation}
In Eq.~(\ref{mutual-nG}), $H_{\rm ps}(a)$ is calculated as
\begin{equation}\label{H(a)}
\begin{array}{l}
H_{\rm ps}(a) = - {\int}{\int} d^2{\alpha}\,\,Q_{\rm out}({\alpha}) \log_2[Q_{\rm out}({\alpha})]
\end{array}
\end{equation}
where $Q_{\rm out}({\alpha})$ is the Q-function of Alice's post-selected state, given by
\begin{equation}\label{Q(alpha)}
\begin{array}{l}
Q_{\rm out}({\alpha}) = {\int}{\int} d^2{\beta}\,\,Q_{\rm out}({\alpha},{\beta}).
\end{array}
\end{equation}
Note that while we can have analytical forms for $Q_{\rm out}({\alpha},{\beta})$ and $Q_{\rm out}({\beta})$, from which we can calculate $H_{\rm ps}(a,b)$ and $H_{\rm ps}(b)$ using Eqs.~(\ref{H(a,b)}) and (\ref{H(b)}), respectively, no closed-form solution for $Q_{\rm out}({\alpha})$ could be used, so Eqs.~(\ref{H(a)}) and (\ref{Q(alpha)}) should be numerically determined.

Now, we calculate the mutual information between Alice and Bob for the parameters of Fig.~{\ref{CM_Inf_cutoff}} using two approaches; first using the covariance matrix, ${{\bf{M_{\rm ps}}}}$, of the amplified (or post-selected) state via Eq.~(\ref{mutual-G}), and also using the Q-function of the post-selected state via Eq.~(\ref{mutual-nG}), with the results shown in Fig.~\ref{mutual-info}. Note that for the numerical integration of Eqs.~(\ref{H(a)}) and (\ref{Q(alpha)}), we divide the integration interval into $m=1000$ equal subintervals. As can be seen from Fig.~\ref{mutual-info}, there is a very small gap between $I_{\rm ps}(a{:}b)$ calculated using the two approaches. More precisely, for $\gamma_c<3$, the mutual information calculated using the covariance matrix, i.e., Eq.~(\ref{mutual-G}) is less than that calculated using the output Q-function, i.e., Eq.~(\ref{mutual-nG}), while for $\gamma_c>3$, the mutual information calculated from Eq.~(\ref{mutual-G}) is higher than that calculated from Eq.~(\ref{mutual-nG}). Note that by increasing the number of sub-intervals, the numerical integration becomes more precise, and the gap becomes smaller. Note also that since the gap between $I_{\rm ps}(a{:}b)$ calculated using the two approaches is very small (less than 0.8 $\%$ even for $m=1000$), for our numerical simulation we have calculated $I_{\rm ps}(a{:}b)$ using the covariance matrix, ${{\bf{M_{\rm ps}}}}$, of the amplified state via Eq.~(\ref{mutual-G}).

\section{Post-selection in the direct reconciliation scenario}\label{AppendixC}

Here, we show the effectiveness of the post-selection in the finite-size regime for the direct reconciliation. Fig.~\ref{PS-DR} shows the achievable secret key rate secure against Gaussian collective attacks in the direct reconciliation scenario as a function of channel distance without the post-selection, and with the Gaussian post-selection (where the cut-off is sufficiently large, i.e., $\gamma_c=3g \sqrt{V_B}$) in the finite-size regime. We found if the block size is sufficiently large, here larger than $n=10^{10}$, the transmission range of the direct reconciliation scheme can be improved by the post-selection, with the improvement increases with increasing the block size. Now, by keeping the block size fixed, we decrease the post-selection cut-off to $\gamma_c=2g \sqrt{V_B}$, where the post-selected data has a non-Gaussian statistics. As it can be seen, this non-Gaussian post-selection is more effective than the Gaussian post-selection, increasing the transmission range from $3.7$ km to $4.7$ km.

\end{document}